\documentclass[final,5p,times,twocolumn]{elsarticle}
\usepackage{amsmath}
\usepackage{amssymb}
\usepackage{graphicx}
\usepackage{color}
\usepackage{hyperref}

\usepackage[mathscr,scaled=1.15]{urwchancal}
\DeclareFontFamily{OT1}{pzc}{}
\DeclareFontShape{OT1}{pzc}{m}{it}%
{<-> s * [1.15] pzcmi7t}{}
\DeclareMathAlphabet{\mathpzc}{OT1}{pzc}{m}{it}

\definecolor{purple}{rgb}{0.5,0,0.5}
\definecolor{blue}{rgb}{0.0,0,0.9}

\biboptions{sort&compress}

\journal{Physics Letters B}

\begin{document}

\begin{frontmatter}

%\title{Extracting the Nakanishi Weight Function for an Euclidean Bethe-Salpeter Amplitude}
\title{A Novel Algorithm for Extracting the Parton Distribution Amplitude from the Euclidean Bethe-Salpeter Wave Function}

\author[PKU1,PKU2]{Fei Gao}
\author[Nankai]{Lei Chang}
\author[PKU1,PKU2,PKU3]{Yu-xin Liu}

\address[Nankai]{School of Physics, Nankai University, Tianjin 300071, China}
\address[PKU1]{Department of Physics and State Key Laboratory of Nuclear Physics and Technology, Peking University, Beijing 100871, China}
\address[PKU2]{Collaborative Innovation Center of Quantum Matter, Beijing 100871, China}
\address[PKU3]{Center for High Energy Physics, Peking University, Beijing 100871, China}

\date{30 Oct 2014}
%\date{21 May 2014}

\begin{abstract}
%
%$\,$\\[-7ex]\hspace*{\fill}{\emph{Preprint no}.}\\[1ex]
%
We propose a new numerical method to compute parton distribution amplitude(PDA) from the Euclidean Bethe-Salpeter wave function.
The essential step is to extract the weight function in the  Nakanishi representation of the Bethe-Salpeter wave function in Euclidean space, which is an ill-posed inversion problem,
 %
 %and can then be solved
 %
 via the maximum entropy method(MEM).
 The Nakanishi weight function as well as the corresponding light-front PDA can be well determined.
 We confirm the previous works on PDA computation therein the different method has been performed.
\end{abstract}

\begin{keyword}
%deep inelastic scattering \sep
%Drell-Yan process \sep
parton distribution function \sep
Nakanishi representation \sep
Bethe-Salpeter amplitude \sep
maximum entropy method \sep
light-front wave function

\smallskip

%\pacs{
%12.38.Aw,   % General properties of QCD (dynamics, confinement, etc.)
%12.38.Lg,   % Other nonperturbative calculations
%11.10.St    %Bound and unstable states; Bethe-Salpeter equations
%%14.40.Be    %Light mesons (S=C=B=0)
%%14.65.Bt,   % Light quarks
%%14.20.Dh,    % Protons and neutrons
%%12.15.-y,    % Electroweak interactions
%}

%\maketitle

\end{keyword}

\end{frontmatter}

\medskip

\noindent\textbf{1.$\;$ Introduction}.
The parton distribution amplitudes (PDAs) of mesons, demonstrating the meson structure on the light front, play an essential role in the hard exclusive processes. The cross sections of the processes can be written as the convolution of the hard-scattering kernel which can be computed perturbatively, and the PDAs of hadron involved~\cite{LB:1980PRD}. The leading twist parton distribution amplitudes of mesons are defined by integrating out the transverse momentum $k_{\perp}^{}$ from the light front wave function, which then can be obtained through projecting meson's Bethe-Salpeter wave function onto the light-front.

Although
%
%there are having some efforts
%
some efforts have been made to calculate Bethe-Salpeter equation(BSE) directly in Minkowski space (see, e.g., Ref.~\cite{Carbonell:2010EPJA}) with the simple scattering kernel, many BSE calculations are still carried out in Euclidean space which are much easier to compute. The challenge in the Euclidean scheme is how to project the discrete Euclidean wave function data on the light-front to get light-front quantities. The Nakanishi representation of wave function provides a natural way to solve this problem. One can transform this challenging question to the proposition that if it is possible to compute weight function of Nakanishi representation when one have an appropriate solution of BSE in Euclidean space.
Nakanishi representation was proposed in Ref.~\cite{Nakanishi:1963PR} to parameterize the relativistic two-particle bound state in Minkowski space.
Although
%
%the
%
there
lacking a proof of uniqueness of the weight function in the nonperturbative case in Euclidean space,
we suppose the wave function can still be parameterized by the similar form in Euclidean space as
\begin{equation}\label{eq:Nakanishi}
\Phi(k,P)=\int^1_{-1}dz\int^\infty_0d\gamma\frac{g(\gamma,z)}{(k^{2}+z k\cdot P+\frac{1}{4}P^{2}+M^{2}+\gamma)^3} \, ,
\end{equation}
where $k^{2}>0$ is the space-like momentum and $P^{2}=-M_{\text{bs}}^{2}$ with $M_{\text{bs}}$ the bound state mass and $M$ is an infrared regulated scale. The weight function $g(\gamma,z)$ is a two-dimensional function in real space. The corresponding leading twist two-particle light-front parton distribution can be defined as
\begin{equation}
\varphi(x)=\int d^{4} k \delta(n\cdot k-xn\cdot P) \Phi(k-\frac{P}{2},P) \, ,
\end{equation}
where $n$ is the light-like vector $n^{2}=0$. We neglect the possible spin structure for simplification at the moment. With the help of $m$-order moment defined as $\int_{0}^{1} dx x^{m}\varphi(x)$,
one can get the one-dimensional integral representation of the light-front PDA in Euclidean space as
\begin{equation}\label{PDAdef}
\varphi(x)=\mathcal{N}\int_{0}^{\infty} d\gamma \frac{g(\gamma,1-2x)}{\gamma+M^{2}-x(1-x)M_{\text{bs}}^{2}} \, ,
\end{equation}
where $\mathcal{N}$ is the corresponding normalization constant to ensure $\int_{0}^{1} dx \varphi(x)=1$.
One can plot the light-front PDA with the complete knowledge of weight function in the Nankanishi representation in hand
%
%, so the
%
. The main aim of this paper is then to provide a practical algorithm
%
%one, i.e., reconstructing
%
to construct the weight function with the numerical data of $\Phi(k,P)$.
We would like
%
%to emphasize
%
to be modest to emphasize that we do not have the ability to prove that
%
%if
%
whether
the weight function is unique definite
%
%from
%
in
a more fundamental view at present.

The techniques for projecting a realistic
%
%pions
%
pion's Bethe-Salpeter (BS) amplitude in Euclidean space onto the
light-front have been pioneered recently~\cite{Chang:2013PRLA}, therein a special Nakanishi representation of each scalar component of the BS amplitude was parameterized to produce the corresponding Euclidean functions and the Mellin moments of PDA can be computed directly,
%
%which was then performed
%
which was implemented
to reconstruct the PDA.
Besides, this Euclidean Nakanishi representation has also been extended to evaluate the pion elastic form factor~\cite{Chang:2013PRLB} and pion valence parton distribution~\cite{Chang:2015PLB}. This technique has been proved successfully in some senses of the integral of $k_{\perp}$ physics(PDA for example), practically at least.
%%
%%, however, the lacking of $\gamma$ dimensional dependence causes the trivial couple of $x$ and %%$k_{\perp}$ in the light-front wave function.
%%Then we have to rely on the complete two-dimensional representation of weight function %%containing interplay between $\gamma$ and $z$ to get the appropriate %light-front wave function.

On the other hand, for the case of heavy meson system, owing to a damping influence from the large quark mass,
%
%the authors in
%
Ref.~\cite{Ding:2016PLB} proposes a `brute-force' approach to
calculate the Mellin moments viz. direct integration using interpolations of the numerical solutions for the quark propagator and Bethe-Salpeter amplitude. A damping factor $1/(1+k^{2}r^{2})^{m}$ has been introduced for each Mellin moment to reduce the oscillation problem in the integration. Such procedure works well for the low
%
%value
%
order
of moments, but uncertainty increases progressively for the higher
%
%moments.
%
order ones.
The limit number of Mellin moments always brings large uncertainty in reconstructing PDA. For example, the PDA near $x=1$ behavior depends on the higher
order
%
%moments
%
moments'
behavior.
%
%The authors in
%
However,
Ref.~\cite{Ding:2016PLB} ignored this issue and supposed
that the
heavy meson PDA exponentially damped
%
%in
%
at
the end-point.

In this letter we provide another numerical method to compute the PDA
that is independent of the approach
given
in Refs.~\cite{Chang:2013PRLA} and \cite{Ding:2016PLB}.
The procedure is straightforward: we first extract the weight function of Nakanishi representation (Eq.~(\ref{eq:Nakanishi})) via
%
%MEM
%
the maximum entropy method (MEM)
and then produce
the
PDA
%
%by
%
with
Eq.~(\ref{PDAdef}).

\medskip

\noindent\textbf{2.$\;$Maximum Entropy Method}. As pointed by the authors
%
%in
%
of
Ref.~\cite{Frederico:2016FBS} that the extracting $g(\gamma,z)$ from
the
$\Phi(k,P)$ is a typical
%
%an
%
ill-posed inversion problem.
They regulate the linear system with a diagonal term $\epsilon I$ and show the unstability of solution with different value of $\epsilon$. They appeal that  more efficient methods are required to obtain a stable and unique solution in the Euclidean space.
In this Letter we will claim
that
the maximum entropy method (MEM) is an appropriate
%
%one
%
algorithm
to solve this problem.
In the following we will firstly illustrate this method by an analytical model of weight function
and then
%
%we will
%
produce
%
%pion and $\eta_c$ PDA respectively.
%
PDAs of $\pi$-meson and $\eta_{c}^{}$-meson.

The MEM~\cite{Bryan:1990EBJ,Asakawa:2000PPNP,Nickel:2007AP} is an approach that can be used to solve an ill-posed inversion problem, in which the number of data points is much smaller than the number of degrees of freedom available to the function whose reconstruction is sought.  Its basis is Bayes' theorem in probability theory~\cite{Jeffreys:1998}, which states the probability of an event ``A'', given that a condition ``B'' is satisfied:
\begin{equation}
P(A|B) = \frac{P(B|A) P(A) }{P(B)}\,,
\end{equation}
where, within the sample space, $P(B|A)$ is the probability that events of type ``A'' satisfy the condition ``B'' (usually denoted as likelihood function); $P(A)$ is the total probability that event ``A'' can occur (commonly referred to prior probability); $P(B)$ is the total probability that condition ``B'' is satisfied (playing the role of normalisation constant).

In making use of the MEM to reconstruct the spectral density function, one works with the conditional probability that a spectral function $g(\gamma,z)$ corresponds to a correlation function $\Phi(k_E)$:
\begin{equation}
P[g | \Phi M ] =\frac{P[\Phi|g M]P[g|M]}{P[\Phi|M]} \, ,
 \end {equation}
%%
%%where $P[\Phi|gM]$, $P[g|M]$ are usually called the likelihood function, the prior probability, %%respectively. $P[\Phi|M]$ is simply a normalization constant independent of $\rho(\omega, \vec{p})$.
%%
%
where $M$ represents the set of all definitions and prior knowledge of the spectral function.

According to the central limit theorem,  the likelihood functional is usually taken as:
\begin{eqnarray}
P[\Phi|g M] & = &\frac{1}{Z_{L}}e^{-L[g]}, \\
L[g] &=& \sum^{N_{data}}_{i}\frac{(\Phi_{data}(k_{i,E}^{})-\Phi_{g}(k_{i,E}^{}))^2}{2\sigma^{2}_{i}}\,,
\end{eqnarray}
where $Z_{L}$ is a normalisation factor; $\{\Phi_{data}(k_{i,E}^{}),i=1,\ldots,N_{data}\}$ are
%
%computed from DSEs;
%
provided with dynamical approaches and $\{\Phi_{\rho}(k_{i,E}^{}),i=1,\ldots,N_{data}\}$ are obtained from Eq.\,\eqref{eq:Nakanishi} using any given model for $g(\gamma,z)$.
One typically chooses $\sigma_{i}^{} = \mathpzc{s}\,\Phi_{data}(k_{i,E})$, with $\mathpzc{s} \lesssim 0.1$.

The central feature of the MEM is the prior probability,
which can be expressed here in terms of the spectral entropy as
\begin{equation}
P[g|M(\alpha,\mathpzc{m})] = \frac{1}{Z_S}e^{\alpha S[g,\mathpzc{m}]} \, ,
\end{equation}
where $Z_{S}$ is a normalisation factor, $\alpha$ is a positive-definite scaling factor, and the exponent involves the Shannon-Jaynes entropy \cite{Shannon:1948BSTJ,Jaynes:1957PRa,Jaynes:1957PRb}
\begin{equation}
S[g,\mathpzc{m}] = \int^1_{-1} dz\int_{0}^{\infty}d\gamma
\Big[ g(\gamma,z)-\mathpzc{m}(\gamma,z)-g(\gamma,z) \log\frac{g(\gamma,z)}{\mathpzc{m}(\gamma,z)} \Big]\,.
\end{equation}
The quantity $\mathpzc{m}(\gamma,z)$ is the ``default model'' of the spectral function, which is usually chosen to be a uniform distribution so as to avoid assumptions about the structure of the spectral density function; viz.,
\begin{equation}
\label{defaultmodel}
\mathpzc{m}(\gamma,z) = \mathpzc{m}_0 \theta(\Lambda - \gamma)\,.
\end{equation}
A MEM result for $g(\gamma,z)$ is considered reliable if it does not depend on the choices for the $\mathpzc{m}_0$ and $\Lambda$.
By adding the entropy functional, the solution of the spectral function $g$ is unique,
at least the most probable.
Thus, with this procedure, people can extract the spectral function from the integration.

\begin{figure}[t] %[hdbp]
\includegraphics[width=0.48\textwidth]{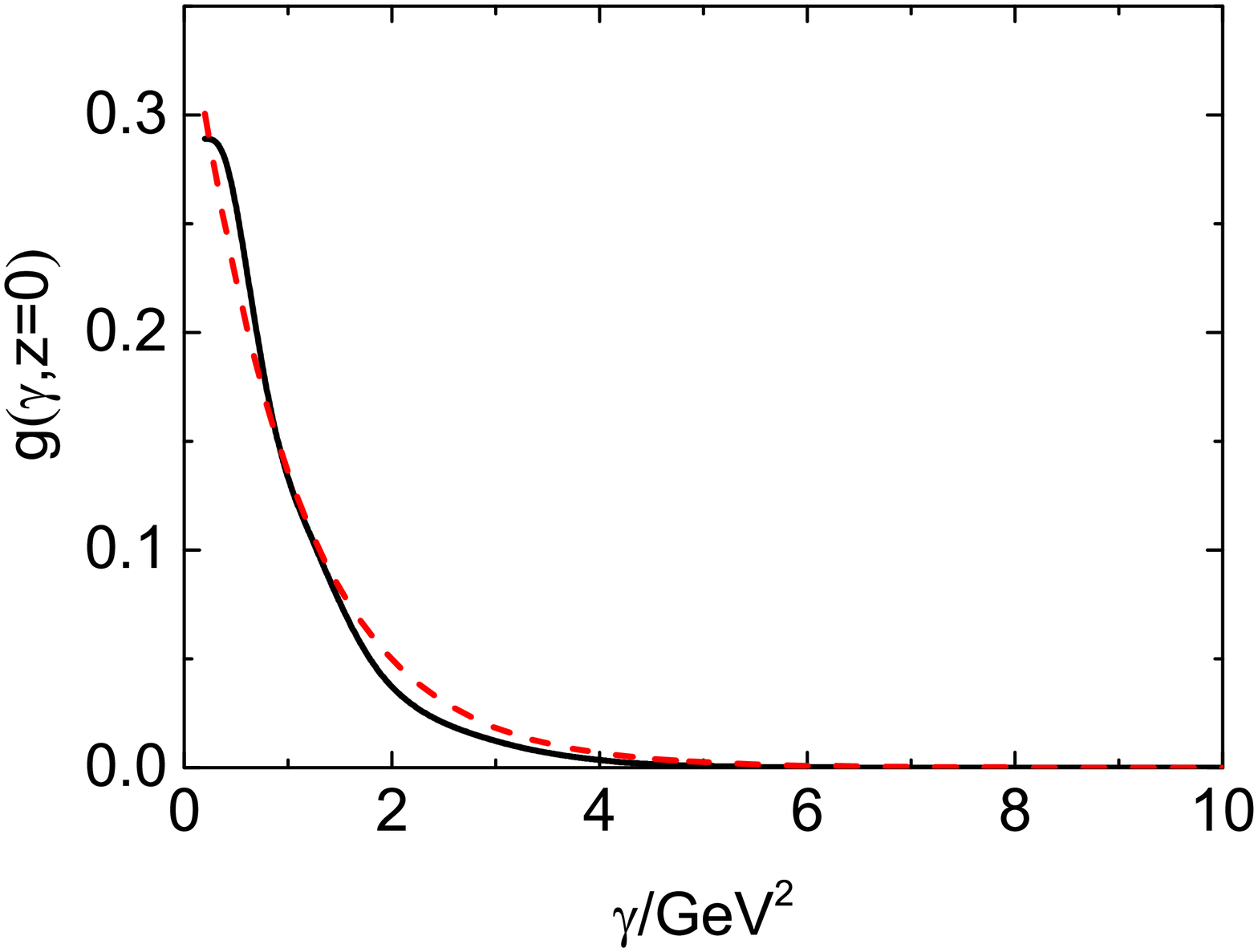}
\includegraphics[width=0.48\textwidth]{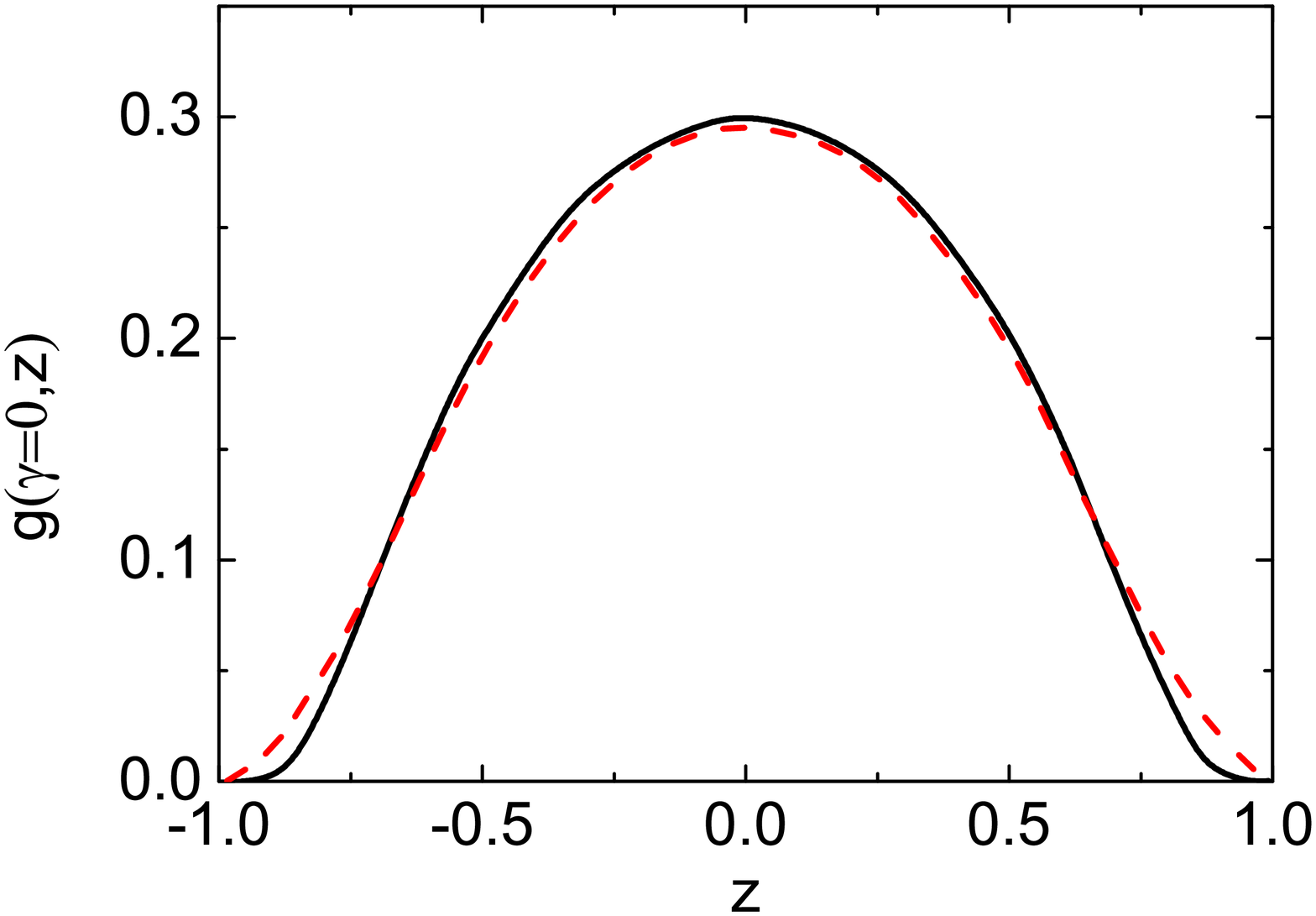}
\caption{Calculated Nakanishi weight function $g(\gamma , z)$ via the MEM and the BSE and DSE (\emph{solid curve}) in cases of $z = 0$ (\emph{upper panel}) and $\gamma = 0$ (\emph{lower panel}) and comparison with the demonstration of the analytical expression in Eq.~(\ref{eqn:ModelinRef8}) (\emph{dashed curve}).
%
%spectral function \emph{solid curve} compared with the real function \emph{dashed %curve}.
}\label{fig:gfun}
\end{figure}

The application of MEM to extract the quark spectral density at finite temperature to explore the properties of strong-interaction matter successfully (see, e.g. Refs.~\cite{Mueller:2010EPJC,Qin:2011PRD,Qin:2013PRD,Gao:2014PRD}) show that
the MEM is a quite promising method to get the spectral density function.
To check further the efficiency of the MEM in exploring the weight function in the  Nakanishi representation of the BS wave function,
we take the Nakanishi weight function model quoted from Ref.~\cite{Frederico:2016FBS}
\begin{equation} \label{eqn:ModelinRef8}
 g(\gamma,z)=e^{-(\gamma+1)/(1-z^2)} \, ,
\end{equation}
with which we can give the variation behavior of the weight function with respect to the parameters $\gamma$ and/or $z$ easily.
We can also get the $g(\gamma , z)$ with the MEM described above where the $\{\Phi_{data}(k_{i,E}^{}),i=1,\ldots,N_{data}\}$ being provided by solving the BSE for mesons and the Dyson-Schwinger equation (DSE) for quark propagator (the calculations will be described in next Section).
We use this model to create the mock data of BS amplitude $\Phi(k_{i},\cos(\theta_{j}))$ with $\cos{\theta}=\frac{k\cdot P}{M_{\text{bs}}\sqrt{k^{2}}}$ by numerically integrating the Nakanishi representation.
The comparison between the numerical result of the Nakanishi weight function we obtained and the demonstration of the above analytical expression is shown in Fig.~\ref{fig:gfun}.
It is evident that the difference between the numerical result and the analytical one is modest
and the point-to-point comparison displays quite good convergence and stability.
We can conclude, by this experience, MEM is an appropriate approach to extract the weight function that can solve the problem
%
%proposed
%
pointed out
in Ref.~\cite{Frederico:2016FBS}.
%%
%%The obtained light front wave function and the real function is shown in Fig.~\ref{fig:pfun}.
%%
%
%\begin{figure}[t] %[hdbp]
%\includegraphics[width=0.48\textwidth]{pgamma.eps}
%\includegraphics[width=0.48\textwidth]{pz.eps}
%\caption{obtained light front wave function\emph{solid curve} compared with the real function\emph{dashed curve}.}\label{fig:pfun}
%\end{figure}

\medskip

\noindent\textbf{3.$\;$ PDAs of $\pi$- and $\eta_{c}^{}$-mesons in Realistic Model}.
The success of extracting the weight function via MEM encourage us to compute PDA from an realistic
BS wave function. In this section we
%
%will use
%
take
the numerical data of pesudoscalar mesons ($\pi$ and $\eta_c$) wave functions
that are calculated within the rainbow-ladder truncation of the BSE and the DSE~\cite{Roberts:1994PPNP,Roberts:2000PPNP,Alkofer:2001PR,Maris:2003IJMPE,Fischer:2006JPG,Bashir:2012CTP,Cloet:2014PPNP} %
to achieve our aim.

The meson wave function will be studied by solving the homogenous BSE in ladder truncation
\begin{eqnarray}\nonumber
\Gamma(k;P) \! & \!\! = \!\! & \! -\frac{4}{3}Z_{2}^{2} \int^{\Lambda}_{d q} \left[\mathcal{G}((k-q)^2)D^f_{\alpha\beta}(k-q) \right. \\\label{eq:homoBSE}
&& \left. \times \, \gamma_{\alpha}^{} S(q_{+})\Gamma(q;P)S(q_{-}) \gamma_{\beta}^{} \right],
\end{eqnarray}
where  $k$ and $P$ are the $q\bar{q}$ state's relative and total momenta, respectively, $q_\pm=q\pm P/2$. The notation $\int^\Lambda_{d q}=\int ^{\Lambda} d^{4} q/(2\pi)^{4}$ stands for a Poincar$\acute{\text{e}}$ invariant regularization of the integral, with $\Lambda$ the regularization mass-scale. The regularization can be removed at the end of all calculations by taking the limit $\Lambda\to\infty$.
$D^f_{\alpha\beta}$ represent the free gluon propagator,
$\mathcal{G}$ denotes the effective interaction and $S$ the quark propagator.
%
%%The most general decomposition of Bethe-Salpeter amplitude for pseudoscalar bound states can be %%written as
%% \begin{equation}
%%  \Gamma^{\text{ps}}(k;P) = \sum_{i=1}^{N^{JP}}\tau_i^{JP}(k;P)\mathcal{F}_i^{JP}(k;P),
%% \end{equation}
%%the index $JP$ represent the angular momentum and the P-parity of the meson, $\tau_i^{JP}(k;P)$ %%are the covariants of the BSA and $\mathcal{F}_i^{JP}%(k;P)$ the Lorentz scalar coefficients.
%%The number of the covariants are $N^{0-}=4$,$N^{0+}=4$ and $N^{2+}=8$. The covariants are %%listed in Appendix. We have considered the charge parity and %all the scalar coefficients are even %%function of the quantity $k\cdot P$.
%%
This equation has solutions at discrete values of $P^{2}=-m_{H}^{2}$, where $m_{H}$ is the meson mass.
The equation also determines completely the BS amplitude $\Gamma(k;P)$ together with the the appropriate normalization condition for the bound states.
The normalization of the meson's BS amplitude is usually taken with the condition~\cite{NakanishiNorm}
\begin{eqnarray}\nonumber
2P_{\mu} \! & \! \! = \!\! & \! \frac{N_c}{N_J}\frac{\partial}{\partial P_\mu}\int^\Lambda_{d q}  \text{tr} \left[ \Gamma(q;-K) \right. \\
&&  \left. \times \, S(q+)\Gamma(q;K)S(q-) \right] \big{|}_{P^2=K^2=-m^2}^{} \, ,
 \end{eqnarray}
where $N_{c}=3$ is the color number and $N_{J}=2J+1$ is the number of the polarization directions of a meson with angular momentum $J$.

The rainbow truncated DSE for the quark propagator in Euclidean space reads
\begin{eqnarray}\nonumber
S(p)^{-1} \! & \!\! = \!\! & \! Z_{2} i\gamma\cdot p + Z_{4} m_{q}(\mu)  + \frac{4}{3}Z^{2}_{2} \int^\Lambda_{d q} \mathcal{G}((p-q)^2) \\
&& \times \, D^{f}_{\alpha\beta}(p-q) \gamma_{\alpha}^{} S(q) \gamma_{\beta}^{} \, ,
\end{eqnarray}
where $Z_{2}$ and $Z_{4}$ are the wave functions and mass renormalization constant, respectively, $m_q(\mu)$ is the current quark mass at space-like renormalization point $\mu$.
We perform a  flavor-independent renormalization scheme as explained in Ref.~\cite{Chang:2015PLB} to define the $Z_{2}$ and $Z_{4}$ at $\mu$.

These equations are consistent and coupled with an effective coupling function $\mathcal{G}(s)$~\cite{Chang:2009PRL},
for which we employ the infrared constant Ansatz~\cite{Qin:2011PRC},
\begin{equation}\label{eq:gluonmodel}
 \frac{\mathcal{G}(s)}{s}=\frac{8\pi^2}{\omega^4}D e^{-s/\omega^2} +\frac{8\pi^2 \gamma_m \mathcal{F}(s)}{\text{ln}[\tau+(1+s/\Lambda^2_{QCD})^2]} \, .
\end{equation}
The first term characterized by the parameters $\omega$ and $D$ determines the intermediate-momentum part of the interaction.
The second term describes the ultraviolet part and produces the correct one-loop perturbative QCD limit.
$\mathcal{F}(s)=[1-exp(-s/[4m_t^2])]/s$, where $m_t=0.5\,$GeV, $\tau=e^2-1$, $\gamma_m=12/(33-2N_{f})$ (usually) with $N_{f}=4$, and $\Lambda^{N_f=4}_{\text{QCD}}=0.234\,$GeV.
The equations are renormalized at the scale $\mu=2.0\,$GeV.

The leading twist PDA of pseudoscalar meson can be defined by
\begin{equation}
{f_{0^{-}}^{}} \varphi(x) = Z_{2}\int_{dk}\delta(n\cdot k - x n\cdot P)
\text{tr}[\gamma_{5}\gamma\cdot n {\chi_{0^{-}}^{}}(k-\frac{P}{2};P)] \, ,
\end{equation}
where we have introduced the BS wave function $\chi_{0^{-}}^{} = S {\Gamma_{0^{-}}^{}} S$
and the lepton decay constant $f_{0^{-}}^{}$.
Two Dirac covariant components contribute to the leading twist PDA, i.e.,
\begin{equation}
\chi(k;P) = \gamma \cdot P {\chi_{1}^{}}(k;P) + \gamma\cdot k {\chi_{2}^{}}(k;P) + \cdots \, ,
\end{equation}
%
%where ${\chi_{1,2}^{}}(k;P)$ can be solved numerically and depend on $(k^{2};k\cdot P)$.
%
where ${\chi_{1,2}^{}}(k;P)$ depend on $(k^{2};k\cdot P)$ and can be determined numerically.
%
%We
%
As usual, we
suppose they can be represented by the Nakanishi form in Eq.~(\ref{eq:Nakanishi}) and the corresponding weight function would be extracted by the MEM.
%
%And then the PDA is computed by Eq.~\ref{PDAdef}.
%
We can then obtain the PDA with Eq.~(\ref{PDAdef}).

\begin{figure}[ht] %[hdbp]
\includegraphics[width=0.48\textwidth]{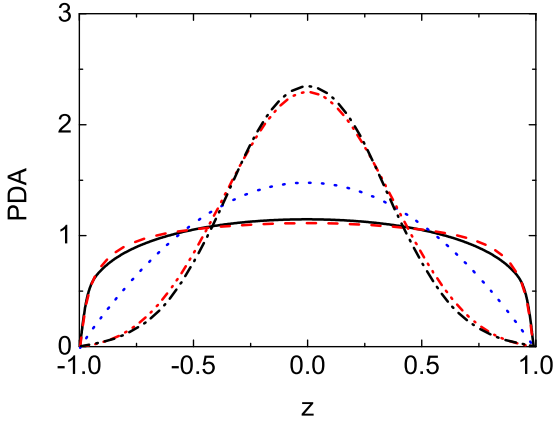}
\caption{Obtained PDA of $\pi $-meson (\emph{solid curve}) and that of $\eta_{c}^{}$-meson (\emph{dot-dashed curve}) and the comparison with previous results ($\phi_{\pi}^{} $ computed by the Nakanishi representation~\cite{Chang:2013PRLA} (\emph{dashed curve}) and $\phi_{\eta_{c}}^{} $ calculated by brute-force procedure~\cite{Ding:2016PLB} (\emph{dot-dot-dashed curve})).
The asymptotic form is also exhibited with dotted curve.}\label{fig:pdamodel}
\end{figure}

We
%
%choose
%
have carried out calculations by choosing the same parameters $\omega=0.5\,\text{GeV}$ and $D\omega=(0.87\,\text{GeV})^{3}$ as the same as those in Ref.~\cite{Chang:2013PRLA} to produce the  $\pi$-meson PDA;
and $\omega=0.8\, \text{GeV}$ and $D\omega=(0.7\,\text{GeV})^{3}$ (the corresponding leptonic decay constant of $\eta_{c}^{}$ is about $0.28\, \text{GeV}$) to calculate the $\eta_{c}^{}$-meson PDA. %
%We plot the $\pi$ and $\eta_c$ PDAs via MEM and their previous results in Fig.~\ref{fig:pdamodel}.
%
The presently obtained results of the $\pi$- and $\eta_{c}^{}$-meson PDAs via MEM and the comparison with previous results are illustrated in Fig.~\ref{fig:pdamodel}.
%

%%Straightforwardly, we obtained PDA for pion with formula $x^{\alpha}(1-x)^{\alpha}$ with %%$\alpha=0.31$ and PDA for $\eta_c$
%%with formula $x(1-x)e^{-a^2(1-2x)^2}$ with $a=1.8$, which are almost the same with the results.
%
%It is remarkable the PDAs via MEM
%
The Fig.~\ref{fig:pdamodel} shows apparently that the PDAs of the pseudoscalar mesons $\pi$ and $\eta_{c}^{}$ presently obtained via the MEM and DSE approach of QCD
match the previous results given in the same dynamical method in the valence region very well
and the slight difference in the middle region of $x$ is tolerable.
%
%The previous results are confirmed
%
Such a good agreement confirms the previous results on one hand and, on the other hand,
indicates that the MEM is efficient to determine the PDA.

\medskip

\noindent\textbf{4.$\;$ Summary and Remarks}.
In this Letter we
%
%consider a pure numerical technique, i.e., MEM
%
propose a practical algorithm to determine the PDA of mesons in the framework of Bethe-Salpeter equation and Dyson-Schwinger equation approach of QCD.
The key point of our new algorithm is implementing the MEM to extract the weight function of
the
Nakanishi representation
%
%and compute the corresponding PDA.
%
of the mesons Bethe-Salpeter wave function.
The merit of the MEM is that one neither needs to rely on the limit knowledge of
the Chebyshev moments of
the mesons'
Bethe-Salpeter amplitude to parameterize the Nakanishi weight function
(like previous $\pi$ case) by special form,
nor has to be restricted by the limit number of Mellin moments
(like previous $\eta_{c}^{}$ case) to suppose some special forms of PDA.
The potential advantage of MEM can be applied to find the light-front wave function of meson
when one has the Bethe-Salpeter wave function in hand,
which we will leave it for future work.
%
%Finally
%
In addition,
we should also admit  some weaknesses of the MEM,
for example, much uncertainty would be encountered when the physical weight function is not positive definite that might be the case of excited state's PDA.
We also run into some difficulty when the Bethe-Salpeter wave function is not monotonous.
However the equivalence of
the
three methods mentioned above
%
%allow
%
allows
us to choose appropriate one to analyze the PDA case-by-case.

%% Ref.\,\cite{Chang:2014Basic}
%%
%%\begin{eqnarray}
%%\label{TqFULL}
%% q(x)&=&N_c  {\rm tr} \int_{dk}^{\Lambda}\!
%%\delta(n\cdot k-x n\cdot P)\,
%%\partial_{k} \left[ \Gamma(k-\frac{P}{2};-P)S(k)\right]\nonumber\\&& \Gamma(k-\frac{P}{2};P) S(k-P)\,,
%%\end{eqnarray}

%%
%%
%%\begin{figure}[t]
%%\centerline{\includegraphics[width=0.8\linewidth]{FigPDFmodelx.PDF}}
%%\caption{\label{fig:PDFmodel} Parton distribution function at $\zeta_{H}$. Curves:
%%\emph{solid},    Rainbow-ladder computation herein corresponding to Eq.(11);
%%\emph{dashed},\, $\rho(z)=\frac{3}{4}(1-z^{2})\left(1+6 a_{2}C_{2}^{3/2}(z)\right)$ with
%%$a_2=0$ (\emph{dotdashed} $a_{2}=2/3$);
%%\emph{dotted}, \,$\rho(z)=\frac{1}{\pi}(1-z^{2})^{-\frac{1}{2}}$. }
%%\end{figure}
%%

%%
%%In Fig.\,\ref{fig:PDFrl}
%%\newpage

\medskip

%\section*{Acknowledgments}
\noindent\textbf{Acknowledgments}.
%
%We are grateful for astute remarks by:.
%
Work supported by the Thousand Talents Plan for Young Professionals (LC),
and the National Natural Science Foundation of China with contract No. 11435001
and the National Key Basic Research Program of China with contract
Nos. G2013CB834400 and 2015CB856900 (FG and YXL).
%%
%%
%%\vspace*{-2ex}

%%\medskip

%%\appendix

%%\setcounter{equation}{0}
%%\setcounter{table}{0}
%%\renewcommand{\theequation}{A\arabic{equation}}
%%\renewcommand{\thetable}{A.\arabic{table}}

%%\noindent\textbf{Appendix}.
%%

%%\smallskip

%%\noindent\textbf{References}.

%%\bibliographystyle{../zchanglei2/zPDA5/model1a-num-names}
%%\bibliography{../../../CollectedBiB}

\begin{thebibliography}{99}
\expandafter\ifx\csname natexlab\endcsname\relax\def\natexlab#1{#1}\fi
\providecommand{\bibinfo}[2]{#2}
\ifx\xfnm\relax \def\xfnm[#1]{\unskip,\space#1}\fi
%Type = Article
%


\bibitem{LB:1980PRD}
     G. P. Lepage, and S. J. Brodsky,
        Phys. Rev. D {\bf 22} (1980),  2157.


\bibitem{Carbonell:2010EPJA}
      J. Carbonell, and V. A. Karmanov,
        Eur. Phys. J. A {\bf 46} (2010), 387.

\bibitem{Nakanishi:1963PR}
      N. Nakanishi,
        Phys. Rev. {\bf 130} (1963), 1230.
%%CITATION = ARXIV:1301.0324;%%

%Type = Article
\bibitem{Chang:2013PRLA}
      L.~Chang, I. C. Cl\"{o}et, J. J. Cobos-Martines, C. D. Roberts, S. M. Schmidt, and P. C. Tandy,
        Phys. Rev. Lett. {\bf 110} (2013), 132001.
%%CITATION = ARXIV:1301.0324;%%

\bibitem{Chang:2013PRLB}
      L.~Chang, I. C. Cl\"{o}et, C. D. Roberts, S. M. Schmidt, and P. C. Tandy,
        Phys. Rev. Lett. {\bf 111} (2013), 141802.
%%CITATION = ARXIV:1301.0324;%%

\bibitem{Chang:2015PLB}
     L.~Chang, and A. W. Thomas,
        Phys. Lett. B {\bf 749} (2015), 547.


\bibitem{Ding:2016PLB}
      M. H. Ding, F. Gao, L. Chang, Y. X. Liu, and C. D. Roberts,
        Phys. Lett. B {\bf 753}  (2016), 330.


\bibitem{Frederico:2016FBS}
      T. Frederico, J. Carbonell, V. Gigante, and V. A. Karmanov,
        Few-Body Syst. {\bf 57} (2016), 549.

\bibitem{Bryan:1990EBJ}
     R.~K. Bryan,
        Eur. Biophys. J. {\bf 18} (1990), 165.

\bibitem{Asakawa:2000PPNP}
     M.~Asakawa, T.~Hatsuda and Y.~Nakahara,
          Prog. Part. Nucl. Phys. {\bf 46} (1990), 459.
%%CITATION = HEP-LAT/0011040;%%

\bibitem{Nickel:2007AP}
      D. Nicke,
         Ann. Phys. {\bf 322} (2007), 1949.

\bibitem{Jeffreys:1998}
     H.~Jeffreys,
\newblock {\em Theory of Probability (Third Edition)} (Oxford University Press,
  Oxford (UK), 1998).

\bibitem{Shannon:1948BSTJ}
     C.~E. Shannon,
       Bell Syst. Tech. J. {\bf 27} (1948), 379.
%%CITATION = BSTJA,27,379;%%

\bibitem{Jaynes:1957PRa}
     E.~T. Jaynes,
         Phys. Rev. {\bf 106} (1957), 620.
%%CITATION = PHRVA,106,620;%%

\bibitem{Jaynes:1957PRb}
     E.~Jaynes,
        Phys. Rev. {\bf 108} (1957), 171.
%%CITATION = PHRVA,108,171;%%

\bibitem{Mueller:2010EPJC}
     J.~A. Mueller, C.~S. Fischer and D.~Nickel,
        Eur. Phys. J. C {\bf 70} (2010), 1037.
%%CITATION = 1009.3762;%%
% Quark spectral properties above Tc from Dyson-Schwinger  equations

\bibitem{Qin:2011PRD}
     S. X. Qin, L. Chang, Y. X. Liu, and C. D. Roberts,
         Phys.\ Rev.\ D {\bf 84} (2011), 014017.
% ``Quark spectral density and a strongly-coupled QGP,''

\bibitem{Qin:2013PRD}
     S. X. Qin, and D. H. Rischke,
         Phys. Rev. D {\bf 88} (2013), 056007.
%Quark Spectral Function and Deconfinement at Nonzero   Temperature

\bibitem{Gao:2014PRD}
     F. Gao, S. X.  Qin, Y. X. Liu, C. D.Roberts, S. M. Schmidt,
       Phys. Rev. D {\bf 89} (2014), 076009.
  %Zero mode in a strongly coupled quark gluon plasma

\bibitem{Roberts:1994PPNP}
     C. D. Roberts and A. G. Williams,
       Prog. Part. Nucl. Phys. {\bf 33} (1994), 477.
% `` Dyson-Schwinger equations and their application to hadronic physics  "

\bibitem{Roberts:2000PPNP}
     C. D. Roberts and S. M. Schmidt,
       Prog. Part. Nucl. Phys. {\bf 45} (2000), S1.
%  " Dyson-Schwinger equations: Density, temperature and continuum strong QCD"

\bibitem{Alkofer:2001PR}
     R. Alkofer,  L. V. Smekal,
       Phys. Rept. {\bf 353} (2001), 281.
% `` The Infrared behavior of QCD Green's functions:  Confinement dynamical symmetry breaking,
%   and hadrons as relativistic bound states " .

\bibitem{Maris:2003IJMPE}
     P. Maris and C. D. Roberts,
       Int. J. Mod. Phys. E {\bf 12} (2003), 297.
%  ``  Dyson-Schwinger equations: A tool for hadron physics "

\bibitem{Fischer:2006JPG}
     C. S. Fischer,
        J. Phys. G {\bf  32} (2006), R253.
%  " Infrared properties of QCD from Dyson-Schwinger equations "

\bibitem{Bashir:2012CTP}
     A. Bashir, L. Chang, I. C. Cloet, B. El-Bennich, Y. X. Liu, C. D. Roberts, and P. C. Tandy,
       Commun. Theor. Phys. {\bf 58} (2012), 79.
% ``  Collective Perspective on Advances in Dyson-Schwinger Equation QCD  ".

\bibitem{Cloet:2014PPNP}
      I. C. Cloet and C. D. Roberts,
        Prog. Part. Nucl. Phys. {\bf 77} (2014), 1.
% ``  Explanation and prediction of observables using continuum strong QCD "


 \bibitem{NakanishiNorm}
       R.E. Cutkosky, and  M. Leon,
           Phys. Rev. {\bf 135} (1964), B1445;
       N. Nakanishi,
           Phys. Rev. {\bf 138} (1965), B1182.

\bibitem{Chang:2009PRL}
      L. Chang, and C. D. Roberts,
         Phys. Rev. Lett. {\bf 103} (2009), 081601.
% `` Sketching the Bethe-Salpeter Kernel " .

\bibitem{Qin:2011PRC}
     S. X. Qin, L. Chang, Y. X. Liu, C. D. Roberts, and D.J. Wilson,
        Phys. Rev. C {\bf 84} (2011), 042202.
% ``Interaction model for the gap equation,''
% doi:10.1103/PhysRevC.84.042202[arXiv:1108.0603 [nucl-th]]
%%CITATION = doi:10.1103/PhysRevC.84.042202;%%
%95 citations counted in INSPIRE as of 25 Sep 2016


\end{thebibliography}

%%\begin{thebibliography}{10}

\end{document}